\documentclass[twocolumn,showpacs,preprintnumbers,amsmath,amssymb,superscriptaddress]{revtex4}
\usepackage{graphicx}
\usepackage{dcolumn}
\usepackage{bm}
\usepackage{longtable}

\usepackage{epsfig}
\usepackage{amssymb}
\usepackage{amsmath}
\usepackage{amsthm}
\usepackage{latexsym}
\usepackage{graphicx}
\def\be{\begin{equation}}
\def\ee{\end{equation}}
\def\bc{\begin{center}}
\def\ec{\end{center}}

\begin{document}

\preprint{APS/123-QED}

\title{Multitasking associative networks}

\author{Elena Agliari}
\affiliation{Dipartimento di Fisica, Universit\`a degli Studi di
Parma, viale Usberti 7/A, 43100 Parma, Italy}
\affiliation{Istituto Nazionale di Fisica Nucleare, Gruppo Collegato di Parma}
\author{Adriano Barra}
\affiliation{Dipartimento di Fisica, Sapienza Universit\`a di
Roma, P.le A. Moro 5, 00182, Roma, Italy}
\author{Andrea Galluzzi}
\affiliation{Dipartimento di Fisica, Sapienza Universit\`a di
Roma, P.le A. Moro 5, 00182, Roma, Italy}
\author{Francesco Guerra}
\affiliation{Dipartimento di Fisica, Sapienza Universit\`a di
Roma, P.le A. Moro 5, 00182, Roma, Italy}
\affiliation{Istituto Nazionale di Fisica Nucleare, Gruppo di
Roma}
\author{Francesco Moauro}
\affiliation{Dipartimento di Fisica, Sapienza Universit\`a di
Roma, P.le A. Moro 5, 00182, Roma, Italy}


\begin{abstract}
We introduce a bipartite, diluted and frustrated, network as a sparse restricted Boltzmann machine and we show
its thermodynamical equivalence to an associative working memory able to retrieve several patterns in parallel without falling into spurious states typical of classical neural networks. We focus on systems processing in parallel a finite (up to logarithmic growth in the volume) amount of patterns, mirroring the low-level storage of standard Amit-Gutfreund-Sompolinsky theory. Results obtained through statistical mechanics, signal-to-noise technique and Monte Carlo simulations are overall in perfect agreement and carry interesting biological insights. Indeed, these associative networks pave new perspectives in the understanding of multitasking features expressed by complex systems, e.g. neural and immune networks.
\end{abstract}

 \maketitle

Neural networks rapidly became the ``harmonic oscillators" of parallel processing: Neurons, thought of as ``binary nodes" (spins) of a network, behave collectively to retrieve information, the latter being spread over the synapses, thought of as the interconnections among nodes. However, common intuition of parallel processing is not only the underlying parallel work performed by neurons to retrieve, say, an image on a book, but rather, for instance, to retrieve the image and, while keeping the book securely in hand, noticing beyond its edges the room where we are reading, still maintaining available resources for further retrieves as a safety mechanism.

Standard Hopfield networks are not able to accomplish this kind of parallel processing \footnote{Apart the network studied in Ref. \cite{amit_a,letizia}, where (a fixed amount of) patterns were contemporary recalled owing to the correlation among them; the network studied in Ref. \cite{8}, where partial inhibition of a single pattern recovery were allowed by chemical modulation; or the approaches with integrate-and-fire neurons of Ref. \cite{hotellone}, which are far away from the associative network framework.}. Indeed, spurious states, conveying corrupted information, cannot be looked at as the contemporary retrieval of several patterns, but they are rather an unwanted outcome, yielding to a glassy blackout \cite{2}.
Such a limit of Hopfield networks can be understood by focusing on the deep connection (in both direct \cite{5} and inverse \cite{bialek2} approach) with restricted Boltzmann machines (RBMs) \cite{17}. In fact, given a machine with its set of visible (neurons) and hidden (training data) units, one gets, under marginalization over the latter, that the thermodynamic evolution of the visible layer is equivalent to that of an Hopfield network. It follows that an underlying fully-connected bipartite RBM necessarily leads to bit strings of length equal to the system size and whose retrieval requires an orchestrated arrangement of the whole set of spins. This implies that no resources are left for further tasks, which is, from a biological point of view, too strong a simplification.

Goal of this paper is to relax this constraint so to extend standard neural networks toward multitasking capabilities, whose interest goes far beyond the artificial intelligence framework  \cite{faye,bear,future}.
In particular, starting from a RBM, we perform dilution on its links in such a way that nodes in the external layer are connected to only a fraction of nodes in the inner layer (fig.$1$, left). As we show, this leads to an associative network which, for non-extreme dilutions, is still embedded in a fully-connected topology, but the bit-strings encoding for information are sparse (i.e. their entries are $+1,-1$ as well as $0$), (fig.$1$, right); for relatively low and large degrees of dilution, this ultimately makes the network able to parallel retrieve without falling into spurious states.

\begin{figure}\label{disegni}
\begin{center}
\includegraphics[width=6.0cm]{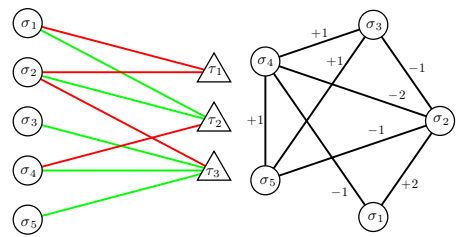}
\caption{(Color online) Example of diluted RBM, whose layers are made up of $N=5$ and $P=3$ elements, (left) and its corresponding weakened associative network (right). In the former brighter (darker) links have positive (negative) coupling; in the latter, the patterns turn out to be $\xi^1=[-1,-1, 0, 0, 0]$, $\xi^2=[+1, +1, 0, -1,0]$, $\xi^3=[0, -1, +1, +1, +1]$ and the weight associated to each link $(i,j)$ is $\sum_{\mu} \xi_i^{\mu}\xi_j^{\mu}$.}
\end{center}
\end{figure}
More precisely, let us denote the $P$ binary spins making up the external layer as $\tau_{\mu} = \pm 1, \mu \in [1,...,P]$ and the $N$ binary spins making up the internal layer as $\sigma_i = \pm 1, i \in [1,...,N]$. RBMs admit the Hamiltonian description
\begin{equation}
H (\sigma, \tau; \xi) = \frac{1}{\sqrt{N}}\sum_{i,\mu}^{N,P}\xi_i^{\mu} \sigma_i \tau_{\mu},
\end{equation}
where we called $\xi_i^{\mu}$ the (quenched) interaction strength between the $i^{th}$ spin of the inner layer and the $\mu^{th}$ spin of the external layer (possibly to be extracted from a proper probability distribution $P(\xi_i^{\mu})$, meant as the outcome of a learning process).
Usually, one defines $\alpha= \lim_{N \to \infty} P/N$ as the storage value; in this work we deal with the ``low storage'' regime, i.e. $P \sim \log N$, corresponding to $\alpha=0$.

The thermodynamics of the system can be obtained by explicit calculation of the (quenched) free energy $f(\beta)$ via the partition function $Z_{N,P}(\beta)$ \cite{22,MPV}, which read off, respectively, as
\begin{eqnarray}
f(\beta) &=& \lim_{N\to \infty}\frac{1}{\beta N} \mathbb{E} \log Z_{N,P}(\beta),\\
Z_{N,P}(\beta) &=& \sum_{\sigma,\tau}\exp\left[-\beta H(\sigma,\tau; \xi)\right],
\end{eqnarray}
$\mathbb{E}$ being the average over the quenched variables $\xi$. A key point here is that the interaction is one-body in each layer, such that marginalizing over one spin variable is straightforward and gives (expanding up to second order the hyperbolic cosine)
\begin{eqnarray}\label{hopfield}
&&Z_{N,P}(\beta) = \sum_{\sigma}\prod_{\mu=1}^P\Big[ \cosh \Big( \frac{\beta}{\sqrt{N}}\sum_i^N \xi_i^{\mu}\sigma_i \Big) \Big]= \\
\label{hopfield2}
&& \sum_{\sigma} e^{\frac{\beta^2}{2N}\sum_{i,j}^{N,N}\sum_{\mu=1}^P \xi_i^{\mu}\xi_j^{\mu} \sigma_i \sigma_j} = \sum_{\sigma}e^{- \frac{N \beta^2}{2} \sum_{\mu=1}^P m_{\mu}^2},
\end{eqnarray}
where we introduced the $P$ Mattis magnetizations $m_{\mu}= N^{-1}\sum_i^N \xi_i^{\mu}\sigma_i$.
When $P(\xi_i^{\mu}=+1)=P(\xi_i^{\mu}=-1)=1/2$, the Hamiltonian implicitly defined in eq. \ref{hopfield2} recovers exactly the Hopfield model (at a rescaled noise level $\beta^2$) and the ansatz of pure state, i.e. $\bold{m}=(1,0,...,0)$ (under permutational invariance) \cite{2}, correctly yields the proper minimization of the free-energy in the low-noise limit.
This means that, once equilibrium is reached, the system configuration $\pmb{\sigma}$ is aligned (under gauge invariance) with pattern $\xi^1$, relaxation that is understood as recovery of a pattern of information.

As anticipated, here we remove the hypothesis of full-connection for the bipartite network, diluting randomly its links in such a way that the coupling distribution gets
\begin{equation}\label{eq:dilution}
P(\xi_i^{\mu})= \frac{1-d}{2} \delta_{\xi_i^{\mu},-1} + \frac{1-d}{2} \delta_{\xi_i^{\mu},+1}+ d \delta_{\xi_i^{\mu},0},
\end{equation}
where $d \in [0,1]$ is a proper dilution parameter and $\delta_{i,j}$ is the Kronecker delta. It is easy to see that with this distribution, after marginalizing over one layer as usual, we get an associative network, where the $P$ patterns $\xi^{\mu}$ ($\mu = 1,...,P$) contain zeros, on average for a fraction $d$ of their length (a sparse coding can also be found in Willshaw's model \cite{willshaw}). As a result, the pure state ansatz can no longer work.
In fact, now, the retrieval of a pattern does not employ all spins and those corresponding to null entries  can be used to recall other patterns. 

\begin{figure}\label{fig:All}
\begin{center}
\includegraphics[width=8cm]{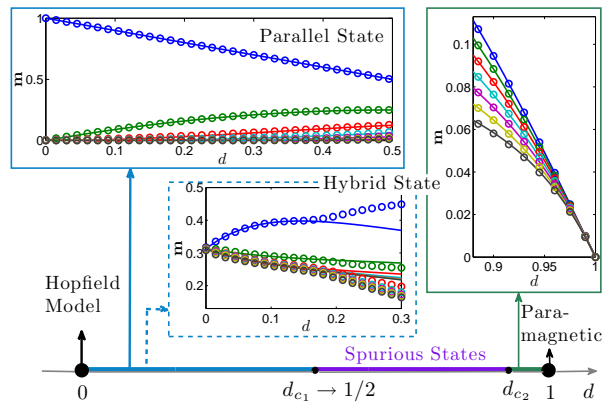}
\caption{(Color on line) Schematic representation of the different regimes exhibited by the systems at $\beta \to \infty$; here we fixed $P=7$, for which $d_{c_1} \approx 0.51$, $d_{c_2} \approx 0.89$. Solid (dashed) line frames denote global (local) minima. The states depicted correspond to eqs.~7 (parallel state) and 11 (hybrid state).}
\end{center}
\end{figure}

In particular, as we will show (see also fig.~$2$), at relatively low degree of dilution ($d<d_{c_1}$), one pattern, say $\mu=1$, is perfectly retrieved, while a fraction $d$ of spins is still available and its overlap with any remaining pattern is, on average, $1-d$; hence, the second best-retrieved pattern, say $\mu=2$, displays a (thermodynamical and quenched) average of the Mattis magnetization equal to $d(1-d)$. Proceeding analogously, one finds 
\be \label{eq:ansatz}
m_k=d^{k-1} (1-d).
\ee
The overall number of retrieved patterns $K$ therefore corresponds to $ \sum_{k=0}^{K-1} (1-d) d^k =1$, with the cut-off at finite $N$ as $(1-d)d^{K-1} \geq N^{-1}$, due to discreteness. For any fixed and finite $d$, this implies $K \lesssim \log N$, which can be thought of as a ``parallel low-storage'' regime of neural networks.
\newline
On the other hand, at larger degrees of dilution ($d > d_{c_1}$) and $P>2$, this state is no longer stable since no magnetization is large enough to yield a field $\xi_i^{\mu} m_{\mu}$ able to align all the related ($\xi_i^{\mu}  \neq 0$) spins; as a result, the system falls into a spurious state where all patterns are partially retrieved, but none exactly. 
Finally, when dilution is extreme ($1-d \sim P^{-1}$), the retrieval of (nearly) all patterns can still be accomplished.
Whenever the global minimum of the system corresponds to the perfect retrieval of at least one pattern, we refer to ``multitasking capabilities'' or, analogously, to ``parallel retrieval''.

Before proceeding with the thermodynamic analysis, we stress that the dilution introduced here is deeply different from the one introduced  early by Sompolinsky \cite{23} or more recently by Coolen et al. \cite{13}, who worked out the Hopfield model embedded in random networks, ranging from Erd\"{o}s-R\'{e}nyi graphs to small-worlds. In those systems, obtained by diluting directly the Hopfield network, the exciting result was the robustness of the (single) retrieval under dilution.
\begin{figure}
\begin{center}
\includegraphics[width=7.5cm, angle=0]{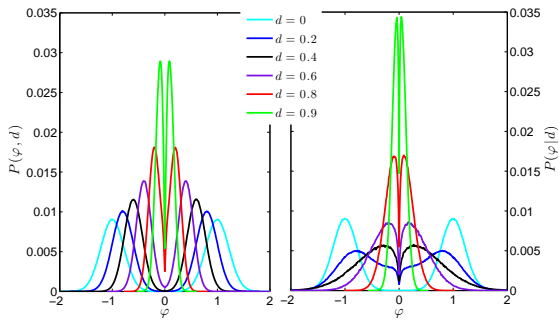}\label{barra}
\caption{Distributions of fields $\varphi$ for dilution \`{a} la Sompolinsky (left) and for our dilution performed on the bipartite network (right), shown for various degrees of dilution, as explained by the legend; for both systems we fixed $\beta \rightarrow \infty$, $N=5000$ and $\alpha=0.05$. In the former case, $d$ represents the average fraction of cut links, in the latter case $d$ represents the average fraction of null pattern entries. As $d$ is tuned, on the left, $P(\varphi|d)$ behaves monotonically corresponding to an Hopfield model embedded on a random graph sparser and sparser, while, on the right, $P(\varphi|d)$ does not behave monotonically and the model is still defined on a fully connected topology.}
\end{center}
\end{figure}

Such different ways of performing dilution - either on links of the associative network (see \cite{23}) or on pattern entries (see Eq.~\ref{eq:dilution}) - yield dramatically different thermodynamic behaviors. To see this let us consider the field insisting on each spin, namely for the generic $i^{th}$ spin $\varphi_i = N^{-1} \sum_{i \neq j=1}^N \sum_{\mu=1}^P \xi_i^{\mu}\xi_j^{\mu}  \sigma_j$, and analyze its distribution $P(\varphi|d)$ at zero noise level. When dilution is realized on links in the direct $\sigma-\sigma$ network (here $d$ is the fraction of links cut), only an average fraction $d$ of the $N$ available spins participates to $\varphi$, in such a way that both the peak and the span of the distribution decrease with $d$ (fig. $2$, left). Conversely, when dilution is realized on links in the underlying bi-layer $\sigma-\tau$ network (here $d$ is the fraction of null entries in a pattern), as $d  >0$, $P(\varphi|d)$ gets broader and peaked at smaller values of fields.
Indeed, at $\beta,\ N$ and $P$ fixed, when dilution is introduced in bit-strings, couplings are made \emph{uniformly} weaker (this effect is analogous to a rise in the fast noise), so that the distribution of spin configurations, and consequently also $P(\varphi| d)$, gets broader.
For small values of $d$ this effect dominates, while at larger values the overall reduction of coupling range prevails and fields get not only smaller but also more peaked  (fig. $2$, right). A topological dilution in the resulting $\sigma -\sigma$ network can be realized also in this case, by taking $d$ sufficiently close to $1$ \cite{future}.

These different scenarios produce different physics, in particular, the latter field distribution can allow parallel retrieval of patterns. The robustness of these multiple basins of attractions can be checked by signal-to-noise analysis \cite{future} and by solving the statistical mechanics of the model as sketched in the following.
We underline that, as no slow noise due to an extensive amount of patterns is at work ($\alpha=0$), replica trick or techniques designed for disordered systems \cite{MPV,22} are not necessary.
We introduce a generic vector for Mattis magnetizations as $\bold{m}=(m_1,...,m_P)$, a density of the states $\mathcal{D}(\bold{m}) = 2^{-N}\sum_{\sigma} \delta(\bold{m}-\bold{m(\sigma)})$ and we write the free-energy density as
\begin{equation}
f(\beta)= \frac{\ln 2}{\beta}+ \frac{1}{\beta N} \log \int d \bold{m} \mathcal{D}(\bold{m})\exp\Big( \frac12 \beta N \bold{m}^2 \Big).
\end{equation}
After introducing the $P$-component vector $\bold{x}$ to allow integral representation of the $P$ delta functions encoded in the density of the states, and after some algebra, this equation becomes
\begin{eqnarray}
f(\bold{m},\bold{x})&=& -\frac{1}{\beta N}\int d \bold{m} d\bold{x} \exp\Big(-N \beta \tilde{f}(\bold{m},\bold{x})\Big), \\
\tilde{f}(\bold{m},\bold{x}) &=& -\frac12 \bold{m}^2 - i \bold{x}\cdot \bold{m} - \frac{1}{\beta}\langle \log 2 \cos[\beta \bold{\xi}\cdot \bold{x}]\rangle_{\xi},
\end{eqnarray}
whose minimization w.r.t. $\bold{m}, \bold{x}$ gives standard saddle-point equation $\bold{m} = \langle \xi \tanh(\beta \bold{\xi}\cdot \bold{m}) \rangle_{\xi}$, whose numerical solution for the case $P=2$ is shown in fig.~$3$. 
When $\beta \to \infty$, stable retrieved states of amplitude $m_1=1-d$ and $m_2=d(1-d)$ are found, in agreement with eq.~\ref{eq:ansatz}.
On the other hand, in the presence of (fast) noise, the dependence on $d$ of the network performance gets more complex.
\begin{figure}\label{fig:mathematica}
\begin{center}
\includegraphics[width=7cm]{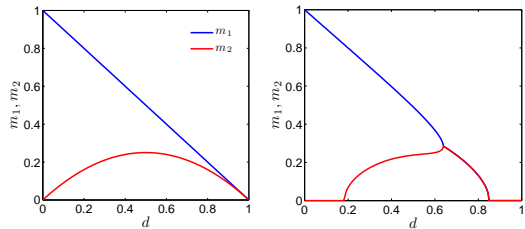}
\caption{(Color online) Two patterns analysis: Analytical solution at $\beta=10^4$ (left) and at $\beta=6.66$ (right). All these curves have been checked versus Monte Carlo simulations (not shown) 
with $N$ up to $10^5$ spins with overall perfect agreement.}
\end{center}
\end{figure}
%
%
%
%
%
%
%
In fact, 
for small $d$, only the first pattern can be retrieved (whenever the fast noise is greater than the signal on $m_2$) and the parallel ansatz $\bold{m}=(d,d(1-d),d(1-d)^2,...)$ recovers the standard pure one (which can be seen as a particular case of the former). This Hopfield-like behavior persists as long as $d(1-d)<\beta^{-1}$, above which $m_2$ also starts to grow and approaches the related zero-noise curve. At intermediate degrees dilution, the two magnetizations $m_1, m_2$ collapse and their amplitude decreases monotonically towards zero.
When $d \sim 1$, the signal on both magnetizations is smaller than fast noise so that retrieval is no longer possible and the system behaves paramagnetically.
We now explain in more detail these features:
We focus on the critical points corresponding to vanishing of magnetizations and to bifurcations, again for the simplest case $P=2$.
The self consistency equations are
\begin{eqnarray}\label{kk}
\nonumber
m_1=d(1-d)\tanh(\beta m_1)+\frac{(1-d)^2}{2}[ \tanh (\beta y)+\tanh(\beta x)] ,\\
\label{kkk}
\nonumber
m_2=d(1-d)\tanh(\beta m_2)+\frac{(1-d)^2}{2}[ \tanh (\beta y ) - \tanh (\beta x)],
\end{eqnarray}
where $y=m_1 + m_2$ and $x= m_1-m_2$.
The critical noise level at which the magnetizations disappear
can be obtained by expanding the self-consistent equation for $m_2$, namely $m_2 \sim (1-d) \beta m_2+\mathcal{O}(m_2^3)$. Therefore, from a standard fluctuation analysis, the critical noise level for the two patterns turns out as $\beta_c=(1-d)^{-1}$, which recovers $\beta_c=1$ for the standard Hopfield model away from saturation \cite{2}.
Critical values of the noise level corresponding to bifurcations can be obtained by expanding for small $x$ and such calculations can be extended to the case $P>2$ (see fig.~$5$); an extensive treatment of the network performances can be found elsewhere \cite{future}.

\begin{figure}
\begin{center}
\includegraphics[width=6.7cm]{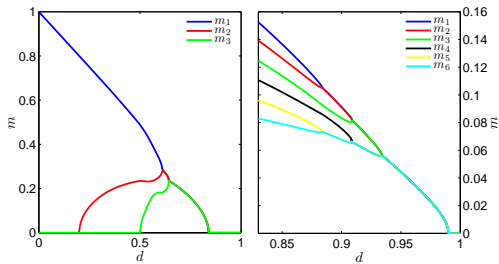}\label{3and6patterns}
\caption{(Color online) Retrieval of three (left) and of six (right) patterns at $\beta=10^2$; for the latter we zoomed on the region of high dilution where bifurcations occur. The number of such bifurcations is, in general, upper bounded by $P-1$.}
\end{center}
\end{figure}

In general, as mentioned above, the case $P>2$ can be much more subtle as, even in the noiseless case, it exhibits several phases (see fig.~2): the parallel ansatz (eq.~\ref{eq:ansatz}) ceases to be stable when $m_1 \leq \sum_{k>1} m_k$, which corresponds to a critical dilution $d_{c_1}$ approaching (exponentially from above) $1/2$ in the limit of large $P$. Within the same region, a ``hybrid state'' $\tilde{s}$, which is a hierachical mixture of all patterns, is also found to be metastable. More precisely, being $\Xi= \sum_{\mu} \xi_i^{\mu}$,
\begin{equation}
{
\tilde{s}_i= (1 - \delta_{\Xi,0})  \textrm{sign}(\Xi) +  \delta_{\Xi,0}[   \xi_i^{1} + \delta_{\xi_i^{1},0} \xi_i^{2} + \delta_{\xi_i^{1},0}\delta_{\xi_i^{2},0} \xi_i^{3} + ...]}.
\end{equation}
This state gets the global minimum whenever $\sum_{i} (1 - \delta_{\Xi,0})  \textrm{sign}(\Xi) \xi_i^{\mu} / N > \sum_{k=1}^{(P-1)/2} \varphi_k (P+1)/(P-k)$, where $\varphi_k=2\sum_l [(1-d)/2]^{2l} d^{P-2l} (P-k)!/[l!(l-1)!(P-k-2l+1)!]$ and $P$ is odd. This condition corresponds to $d>d_{c_2}$, where $d_{c_2}$ converges to $1$ as $P$ gets larger. 

As a final remark, we underline that, although the steady state of the current model and an arbitrary spurious state both display non-zero overlap with several patterns, they are still deeply different. In particular, here the retrieval of multiple patterns corresponds to absolute energy minima (in the noiseless case this holds for any $d>0$) and at least one pattern is exactly retrieved. However, the present model is not devoid of genuine spurious-states, which are, in general, mixtures of all patterns. These states can be destabilized by decreasing $\beta$ (analogously to the standard Hopfield model) or, interestingly, by either increasing or decreasing $d$. 

In summary, the structural equivalence of associative networks and RBMs allows significant developments, both practical and theoretical.
For instance, one can simulate the dynamics of these networks by dealing with an update of $N+P$ spins and a storage of only $NP$ synapses, instead of updating $N$ spins and storing $\sim N^2$ synapses. Moreover, the equivalence suggests that traditional associative networks, where the whole set of neurons needs to be properly arranged in order to achieve retrieval, are not optimal. We overcome this constraint by diluting the links of the RBM, which translates into partially blank patterns. Interestingly, the resulting associative network is not only still able to perform retrieval, but it can actually retrieve several patterns contemporary, without falling into spurious states.
This is an important step toward real autonomous parallel processing and may find applications not only in artificial intelligence \cite{faye}, but also in biological contexts \cite{bear}. For instance, when applied to the modeling of the adaptive immune system, this result allows to see that the (lymphocyte) network is able to successfully respond to several pathogens at once \cite{future}.



The FIRB grant RBFR08EKEV, Sapienza University and INFN are acknowledged for financial support.

\end{document}